\documentclass[12pt]{article}
\usepackage{graphicx}
 \usepackage{cite}

\newcommand\pubnumber{SNSN-XXX-YY}
\newcommand\pubdate{\today}

\def\Title#1{\begin{center} {\Large #1 } \end{center}}
\def\Author#1{\begin{center}{ \sc #1} \end{center}}
\def\Address#1{\begin{center}{ \it #1} \end{center}}

\newcommand\pubblock{\rightline{\begin{tabular}{l} \pubnumber\\
         \pubdate  \end{tabular}}}
\newenvironment{Abstract}{\begin{quotation}  }{\end{quotation}}
\newenvironment{Presented}{\begin{quotation} \begin{center} 
             PRESENTED AT\end{center}\bigskip 
      \begin{center}\begin{large}}{\end{large}\end{center} \end{quotation}}

\begin{document}
\begin{titlepage}
\pubblock

\vfill
\Title{ Proton Charge Radius and Precision Tests of QED }
\vfill
\Author{Jan C. Bernauer}
\Address{Laboratory for Nuclear Science\\ Massachusetts Institute of
  Technology, 77 Massachusetts Ave, Cambridge, MA 02139}
\vfill
\begin{Abstract}
The ``proton radius puzzle'' remains unsolved since it was established in 2010.
This paper summarizes the current state and gives an overview over
upcoming experiments. 
\end{Abstract}
\vfill
\begin{Presented}
XXXIV Physics in Collision Symposium \\
Bloomington, Indiana,  September 16--20, 2014
\end{Presented}
\vfill
\end{titlepage}
\def\thefootnote{\fnsymbol{footnote}}
\setcounter{footnote}{0}

\section{The proton radius puzzle}
The properties of the proton, one of the basic building blocks of the
matter around us, have been a research target for a long
time. Recently, a series of precise experiments of the proton’s charge
radius have produced results which are in strong disagreement, casting
doubt on our knowledge of one of the proton’s fundamental properties
and our understanding of the underlying physics. The discrepancy falls
between the different methods for measuring the proton’s radius. The
following sections address the three principal methods employed to
date.

\subsection{Elastic electron-proton scattering}
The cross section for scattering an electron beam off a proton target
in the first Born approximation is given by
\begin{equation}
\frac{d\sigma}{d\Omega}=\left(\frac{d\sigma}{d\Omega}\right)_\mathsf{Mott}\frac{1}{\epsilon(1+\tau)}\left[\epsilon
Q_E^2\left(Q^2\right)+\tau G_M^2\left(Q^2\right)\right],
\end{equation}
with the negative four-momentum-transfer $Q^2$, the kinematical
variables $\tau=Q^2/4m_p^2$,
$\epsilon=1/(1+2(1+\tau)\tan^2\frac{\theta}{2})$, the Mott cross section
$(d\sigma/d\Omega)_\mathsf{Mott}$ and the electric and magnetic form
factors $G_E$,$G_M$.
Exploiting the cross section's dependency on the kinematical variables, one can
disentangle both form factors from a series of cross section
measurements, for example via the Rosenbluth separation method. This
gives access to the form factors over a large range of four-momentum
transfers.

The root-mean-square charge radius, $r_e$, is defined in terms of the
slope of the electric form factor $G_E$ at $Q^2=0$,
\begin{equation}
r_e=6\hbar^2\left.\frac{dG_E}{dQ^2}\right|_{Q^2=0}.
\end{equation} 

Extractions of this type typically achieve uncertainties on the order
of 1\%.

\subsection{Hydrogen spectroscopy}
The finite size of the proton shifts the atomic energy levels of
hydrogen by small amounts. This effect can be measured in the lower
S-states. Historically, the proton radius has been a correction to level calculations in high
precision QED tests. Experiments have progressed to a state where the uncertainty in the radius is now one of the
limiting factors. Turning the argument around and assuming the
correctness of QED, these kind of experiments can be used to extract a
proton radius.

One way is to measure the 2S-2P transition, which gives the
Lamb shift, and with that the proton radius. 
In a different approach, the transition 1S-2S and a second transition
from 2S to a higher state like 8S or 8D are measured. The proton
radius is then extracted using simultaneous fit of the radius and the
Rydberg constant. 

Currently published measurements are typically less precise than the
results from scattering experiments. Combining the available data,
however, leads to an extraction with similar uncertainties.

\subsection{Muonic Hydrogen spectroscopy}
In recent years, it became possible to study muonic hydrogen, the bound state of a proton and a muon, with
spectroscopy. The muon, because of its larger mass, has a $200$ times
smaller orbit and with that an about $200^3$ higher probability of being
inside the proton. Consequently, the finite size effect is
substantially larger, making a more precise extraction of the
radius possible. The published results quote more than
ten times smaller uncertainties.

\subsection{The puzzle}
\begin{figure}
\centering
\includegraphics{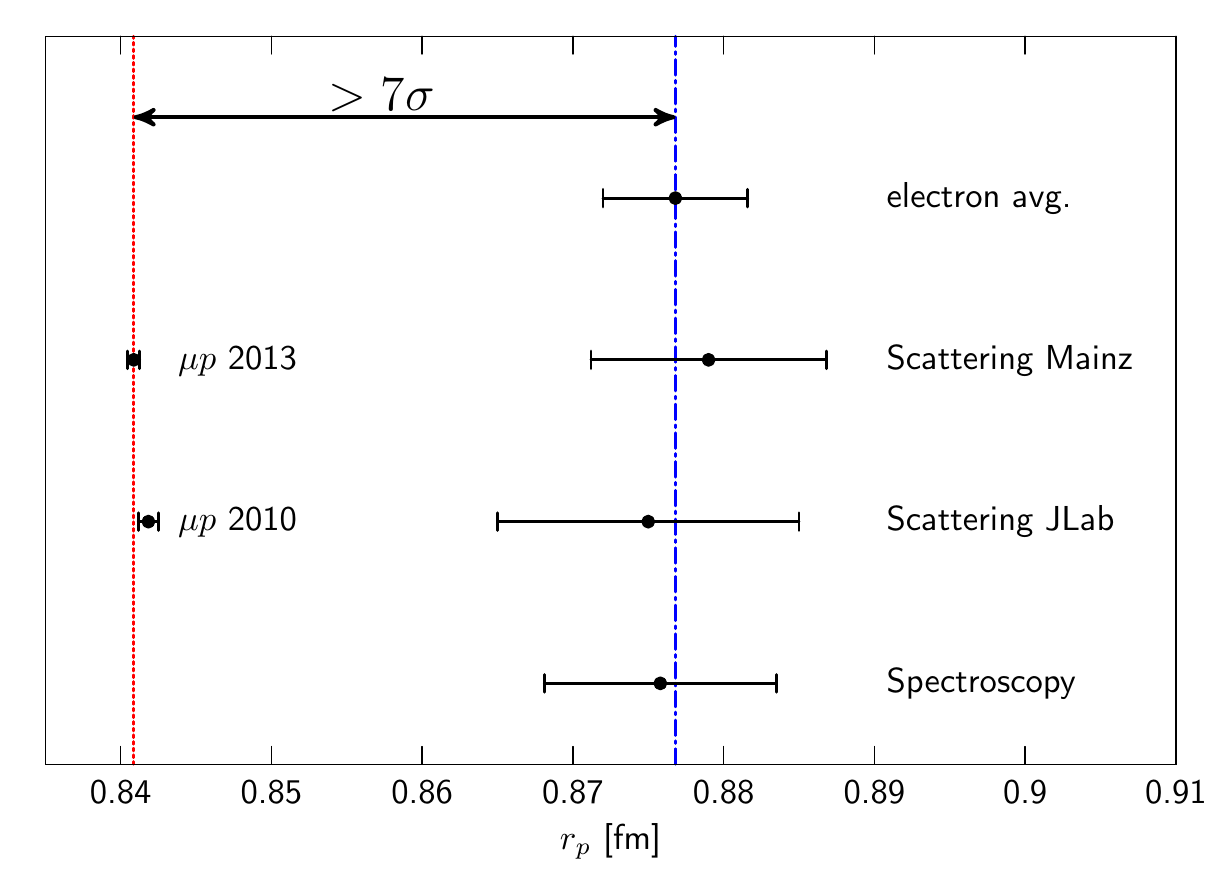}
\caption{\label{radii} Selected results for the proton radius. The
  extractions from electronic measurements from Mainz
  \cite{Bernauer:2010wm,Bernauer:2013tpr} and Jefferson Lab
  \cite{Ron:2011rd} are in agreement with spectroscopy results
  \cite{Mohr:2012tt}, but in strong disagreement with the muonic
  results \cite{Antognini:1900ns,Pohl:2010zza}.}
\end{figure}

Figure \ref{radii} shows the result of recent determinations. For
scattering, the results from \cite{Bernauer:2010wm,Bernauer:2013tpr} and \cite{Ron:2011rd} are
presented. The former is the result of a measurement of more than 1400
cross sections, about twice of all other existing proton form factor
data. The latter is an extraction using almost all available data
except the Mainz data set. It is therefore independent. The
H-spectroscopy result is taken from the global fit of CODATA
2010 \cite{Mohr:2012tt}. These measurements are all in agreement with each other; a
combined result of the electron measurements is shown as ``electron
avg''.  In contrast to this, the two published results from muon
spectroscopy \cite{Antognini:1900ns,Pohl:2010zza}, are consistent
with each other, but more than 7 standard
deviations away from the electron result.

The discrepancy, dubbed the ``proton radius puzzle'', has driven a
wide range of theoretical and experimental efforts, and has even found
its way into popular science literature \cite{Bernauer:2014cwa}. Since
the publication of \cite{Bernauer:2010wm} and \cite{Pohl:2010zza} in
2010, it has withstood all attempts at a solution.

\section{Possible solutions}
The puzzle has prompted a lot of research, leading to a large number of
papers trying to solve the discrepancy. However, many have been ruled
out by further studies. None has seen widespread acceptance by the
community so far. In this section, some of the proposed
solutions will be discussed. Due to the sheer number, this paper can only highlight some of the
ideas. Instead, I will try to give a categorization and a personal
perspective.

\subsection{Errors in experiment execution}
Errors in execution of either the muon experiment, or both atomic hydrogen
spectroscopy and scattering could explain the difference. 
On the muon spectroscopy side, the measured resonance is shifted from
the range expected by the electronic results by far more than its
width, and the results from both measured lines are very
consistent. Only a few concerns about possible problems have been
raised, all of which have been studied and ruled out by the CREMA
collaboration.

On the other hand, conspiring mistakes in the electron-based
extractions seem unlikely, just by the number of experiments which are
all in agreement. However, it is worthwhile to note that the bulk of
spectroscopy results are produced by the same group and a systematic
error might affect all results at once. 

The different data sets produced in scattering experiments by
different groups with different setups span decades in time. It is
remarkable how well they are in agreement \cite{Bernauer:2013tpr}. Specific concerns regarding the Mainz
experiment have been subsequently ruled out by the Mainz collaboration.

\subsection{Errors in experiment analysis}
For the electron experiments, errors in analysis 
procedures are a greater concern, since they may affect multiple experiments in the same way.
For scattering experiments, the extrapolation of the data to $Q^2=0$ is
crucial. While the possibility of structure below the Mainz data set seems
unlikely---structure there would mean that there is a surprising large
amount of the charge at very large radii---it may be possible to find
form factor models which do not exhibit this problem, produce a small
radius and still fit the experimental data well. Many fits of
different parts of the world data set by different groups with
different models have been performed. On the one hand, most of them reproduce the large
radii, with a tendency to be even larger ({\em e.g.}\
\cite{Blunden:2005jv,Borisyuk:2009mg,Graczyk:2014lba,Hill:2010yb,Sick:2005az}). On the
other hand, the fits in
\cite{Bauer:2012pv,Belushkin:2006qa,Hammer:2003ai,Lorenz:2012tm} produce a radius
compatible with the muon result, although with a substantially larger $\chi^2/d.o.f.$.

The scattering experiments have a rather small dependence on
theoretical corrections, which are mostly well understood. However,
several papers ({\em e.g.}\
\cite{Arrington:2011kv,Bernauer:2011zza,Arrington:2012dq})
investigated the effect of different treatments of
two-photon-exchange, {\em i.e.}\ higher Born terms. They tend to
reduce the radius, but can not explain the full discrepancy.

The spectroscopy results rely on theory to a much larger extent. In
the wake of the puzzle, all components have been rechecked and improved, without
finding relevant changes. One remaining uncertainty stems from the
proton polarizability (see \cite{Miller:2011yw} and its citations),
however, the general consensus seems to be that the effect is too
small to explain the discrepancy.

\subsection{New physics}
The theory calculations rely on the assumption that our current
understanding of the underlying physics is correct. If the puzzle
survives any attempts to solve it within the standard theoretical framework, the
solution might present itself in the most welcome outcome: new
physics. Several such solutions have been proposed, and they can
roughly be divided into two groups.

The first group introduces new concepts which invalidate some of the
assumptions, but without any or with little modification to the
Standard Model. This includes papers like \cite{walcher}, in which
Walcher criticizes the approach in the theoretical calculations, and 
\cite{jentschura}, in which Jentschura proposes the existance of
light sea fermions.

The second group introduces bigger changes to the Standard
Model. Particular interesting is the introduction of light dark
matter (see, {\em e.g.}, \cite{Barger:2010aj,Batell:2011qq,Carlson:2012pc,Carlson:2013mya,Karshenboim:2014tka,TuckerSmith:2010ra}) since it might simultaneously solve the muon g-2
anomaly. However, it seems that simple models for such dark matter can
not reconcile both, and in fact, need a lot of fine-tuning to avoid being
ruled out by other experiments. Li, Chen, Wang and Ni \cite{Li:2013dwa,Wang:2013fma} propose large extra dimensions to solve the puzzle.

\section{Upcoming experiments}
To study the puzzle further, a multitude of experiments are under way
or planned. Indeed it is the expressed opinion of the majority of the community that
such new data are needed to make any headway towards a solution.

\subsection{Spectroscopy}
On the muonic side, the CREMA collaboration has switched gears and is studying the
spectra of heavier nuclei, with experiments on deuterium, $^3$He
and $^4$He currently under analysis. Preliminary results from
muonic deuterium indicate that the isotope shift, the difference between
the proton and deuterium radius, is in good agreement with the value
determined from electronic systems. The group also indicated that for
helium, the radii are not different from the electronic values,
however there is still a large theoretical uncertainty.

Several groups at LKB, MPQ, NPL and York University are trying to improve on the existing electronic
hydrogen measurements using a variety of techniques. 

Furthermore, groups at MPQ and VU University Amsterdam are preparing experiments on helium
ions, while NIST is focusing on highly charged ions.

\subsection{Scattering experiments}
The A1 collaboration in Mainz, the same group that performed the high
precision proton form factor measurement, is pursuing several
experiments regarding the radius. With an approach similar to the proton form factor measurement, data
were taken on electron-deuterium scattering in 2012. The analysis is
ongoing. It is planned to extend the proton measurement to higher
four-momenta using the higher beam energies of MAMI-C.

As described earlier, the extrapolation to $Q^2=0$ is critical. Using
initial state radiation to lower the effective beam energy, Mainz aims
to lower the current limit for $Q^2_\mathsf{min}$ by more than an
order of magnitude \cite{Mihovilovic:2014nta}. The approach has significantly different
systematics than the classical approach. A confirmation of the radius
will therefore rule out several possible sources of error.
Data were taken in 2013, and the data analysis is ongoing.

With a similar aim, but different methodology, is the PRad experiment \cite{Gasparian:2014rna} at
Jefferson Lab. Using a high resolution, large acceptance hybrid
calorimeter and a windowless target, elastic scattering at extremely
forward angles will be studied. The concurrent measurement of M\o ller
scattering will enable good control of the absolute normalization.

A missing piece in the puzzle will be filled by the MUSE
collaboration: While we have results from both scattering and
spectroscopy on the electronic side, we miss precise scattering data
of muons off of protons. The MUon Scattering Experiment
\cite{Gilman:2013eiv} aims to measure both $e^{+/-}$ and $\mu^{+/-}$
cross sections using the powerful low-energy beam of the Paul Scherrer
Institute. The measurement of the charge symmetry will allow the group
to control two-photon exchange corrections, while the measurement of
both species with the same detector will enable a direct comparison of
the cross sections and radii, with small systematic uncertainties.

\section{Conclusion}
In the four years after its discovery, the proton radius puzzle has not
been solved. If anything, it has grown stronger. In the next five years, a large number of experiments will
shed some more light on this intriguing problem.


\begin{thebibliography}{99}



\bibitem{Bernauer:2010wm} 
  J.~C.~Bernauer {\it et al.}  [A1 Collaboration],
  Phys.\ Rev.\ Lett.\  {\bf 105}, 242001 (2010)
  [arXiv:1007.5076 [nucl-ex]].


\bibitem{Bernauer:2013tpr} 
  J.~C.~Bernauer {\it et al.}  [A1 Collaboration],
  Phys.\ Rev.\ C {\bf 90}, 015206 (2014)
  [arXiv:1307.6227 [nucl-ex]].

\bibitem{Ron:2011rd} 
  G.~Ron {\it et al.}  [Jefferson Lab Hall A Collaboration],
  Phys.\ Rev.\ C {\bf 84}, 055204 (2011)
  [arXiv:1103.5784 [nucl-ex]].

\bibitem{Mohr:2012tt} 
  P.~J.~Mohr, B.~N.~Taylor and D.~B.~Newell,
  Rev.\ Mod.\ Phys.\  {\bf 84}, 1527 (2012)
  [arXiv:1203.5425 [physics.atom-ph]].
 

\bibitem{Antognini:1900ns} 
  A.~Antognini, F.~Nez, K.~Schuhmann, F.~D.~Amaro, F.~Biraben, J.~M.~R.~Cardoso, D.~S.~Covita and A.~Dax {\it et al.},
  Science {\bf 339}, 417 (2013).

\bibitem{Pohl:2010zza} 
  R.~Pohl, A.~Antognini, F.~Nez, F.~D.~Amaro, F.~Biraben, J.~M.~R.~Cardoso, D.~S.~Covita and A.~Dax {\it et al.},
  Nature {\bf 466}, 213 (2010).

\bibitem{Bernauer:2014cwa} 
  J.~C.~Bernauer and R.~Pohl,
  Sci.\ Am.\  {\bf 310}, no. 2, 18 (2014).


\bibitem{Blunden:2005jv} 
  P.~G.~Blunden and I.~Sick,
  Phys.\ Rev.\ C {\bf 72}, 057601 (2005)
  [nucl-th/0508037].

\bibitem{Borisyuk:2009mg} 
  D.~Borisyuk,
  Nucl.\ Phys.\ A {\bf 843}, 59 (2010)
  [arXiv:0911.4091 [hep-ph]].

\bibitem{Graczyk:2014lba} 
  K.~M.~Graczyk and C.~Juszczak,
  arXiv:1408.0150 [hep-ph].

\bibitem{Hill:2010yb} 
  R.~J.~Hill and G.~Paz,
  Phys.\ Rev.\ D {\bf 82}, 113005 (2010)
  [arXiv:1008.4619 [hep-ph]].

\bibitem{Sick:2005az} 
  I.~Sick,
  Prog.\ Part.\ Nucl.\ Phys.\  {\bf 55}, 440 (2005).

\bibitem{Bauer:2012pv} 
  T.~Bauer, J.~C.~Bernauer and S.~Scherer,
  Phys.\ Rev.\ C {\bf 86}, 065206 (2012)
  [arXiv:1209.3872 [nucl-th]].


\bibitem{Belushkin:2006qa} 
  M.~A.~Belushkin, H.-W.~Hammer and U.-G.~Meissner,
  Phys.\ Rev.\ C {\bf 75}, 035202 (2007)
  [hep-ph/0608337].


\bibitem{Hammer:2003ai} 
  H.~W.~Hammer and U.~G.~Meissner,
  Eur.\ Phys.\ J.\ A {\bf 20}, 469 (2004)
  [hep-ph/0312081].


\bibitem{Lorenz:2012tm} 
  I.~T.~Lorenz, H.-W.~Hammer and U.~G.~Meissner,
  Eur.\ Phys.\ J.\ A {\bf 48}, 151 (2012)
  [arXiv:1205.6628 [hep-ph]].



\bibitem{Arrington:2011kv} 
  J.~Arrington,
  Phys.\ Rev.\ Lett.\  {\bf 107}, 119101 (2011)
  [arXiv:1108.3058 [nucl-ex]].

\bibitem{Bernauer:2011zza} 
  J.~C.~Bernauer, P.~Achenbach, C.~Ayerbe Gayoso, R.~Bohm, D.~Bosnar, L.~Debenjak, M.~O.~Distler and L.~Doria {\it et al.},
  Phys.\ Rev.\ Lett.\  {\bf 107}, 119102 (2011).

\bibitem{Arrington:2012dq} 
  J.~Arrington,
  J.\ Phys.\ G {\bf 40}, 115003 (2013)
  [arXiv:1210.2677 [nucl-ex]].

\bibitem{Miller:2011yw} 
  G.~A.~Miller, A.~W.~Thomas, J.~D.~Carroll and J.~Rafelski,
  Phys.\ Rev.\ A {\bf 84}, 020101 (2011)
  [arXiv:1101.4073 [physics.atom-ph]].

\bibitem{walcher}
Th.~Walcher, [arXiv:1207.4901 [physics.atom-ph]].

\bibitem{jentschura}
U.~D.~Jentschura,
Phys.\ Rev.\ A {\bf 88}, 062514 (2013) 
[arXiv:1401.3666 [physics.atom-ph]].

\bibitem{Barger:2010aj} 
  V.~Barger, C.~W.~Chiang, W.~Y.~Keung and D.~Marfatia,
  Phys.\ Rev.\ Lett.\  {\bf 106}, 153001 (2011)
  [arXiv:1011.3519 [hep-ph]].

\bibitem{Batell:2011qq} 
  B.~Batell, D.~McKeen and M.~Pospelov,
  Phys.\ Rev.\ Lett.\  {\bf 107}, 011803 (2011)
  [arXiv:1103.0721 [hep-ph]].

\bibitem{Carlson:2012pc} 
  C.~E.~Carlson and B.~C.~Rislow,
  Phys.\ Rev.\ D {\bf 86}, 035013 (2012)
  [arXiv:1206.3587 [hep-ph]].

\bibitem{Carlson:2013mya} 
  C.~E.~Carlson and B.~C.~Rislow,
  Phys.\ Rev.\ D {\bf 89}, no. 3, 035003 (2014)
  [arXiv:1310.2786 [hep-ph]].

\bibitem{Karshenboim:2014tka} 
  S.~G.~Karshenboim, D.~McKeen and M.~Pospelov,
  Phys.\ Rev.\ D {\bf 90}, 073004 (2014)
  [Addendum-ibid.\ D {\bf 90}, no. 7, 079905 (2014)]
  [arXiv:1401.6154 [hep-ph]].

\bibitem{TuckerSmith:2010ra} 
  D.~Tucker-Smith and I.~Yavin,
  Phys.\ Rev.\ D {\bf 83}, 101702 (2011)
  [arXiv:1011.4922 [hep-ph]].

\bibitem{Li:2013dwa} 
  Z.~Li and X.~Chen,
  arXiv:1303.5146 [hep-ph].

\bibitem{Wang:2013fma} 
  L.~B.~Wang and W.~T.~Ni,
  Mod.\ Phys.\ Lett.\ A {\bf 28}, 1350094 (2013)
  [arXiv:1303.4885 [hep-ph]].

\bibitem{Mihovilovic:2014nta} 
  M.~Mihovilovic, H.~Merkel and A.~Weber,
  PoS Bormio {\bf 2014}, 051 (2014).

\bibitem{Gasparian:2014rna} 
  A.~Gasparian [PRad at JLab Collaboration],
  EPJ Web Conf.\  {\bf 73}, 07006 (2014).

\bibitem{Gilman:2013eiv} 
  R.~Gilman {\it et al.}  [MUSE Collaboration],
  arXiv:1303.2160 [nucl-ex].
\end{thebibliography}
\end{document}